# Valence band, electron correlation, and possible temperature dependent change in the electronic structure near $E_F$ of a candidate of ambient-pressure nickelate superconductors $Sr_3Ni_2O_5Cl_2$

Ryota Goban[1], Daiki Ootsuki[1,2], Kenichi Ozawa[3], Akira Yasui[4], Masayuki Ochi[5,6,7], Kazuhiko Kuroki[6], Ryo Matsumoto[8], Kensei Terashima[8], Yoshihiko Takano[8], Yuji Muraoka[1,2], and Takayoshi Yokoya[1,2*]

[1]*Graduate School of Environmental, Life, Natural Science and Technology, Okayama University, 1-1-1 Tsushima-naka, Okayama 700-8530, Japan*

[2]*Research Institute for Interdisciplinary Science, Okayama University, 1-1-1 Tsushima-naka, Okayama, 700-8530, Japan*

[3]*Institute of Materials Structure Science (IMSS), KEK, Photon Factory, 1-1 Oho, Tsukuba, Ibaraki 305-0801, Japan*

[4]*Japan Synchrotron Radiation Research Institute (JASRI), 1-1-1 Koto, Sayo, Hyogo 6795198, Japan*

[5]*Forefront Research Center, The University of Osaka, 1-1 Machikaneyama-cho, Toyonaka, Osaka 560-0043, Japan*

[6]*Department of Physics, The University of Osaka, 1-1 Machikaneyama-cho, Toyonaka, Osaka 560-0043, Japan*

[7]*RIKEN Center for Computational Science, 2-1 Hirosawa, Wako, Saitama 351-0198, Japan*

[8]*International Center for Materials Nanoarchitectonics (MANA), National Institute for Materials Science, 1-2-1 Sengen, Tsukuba, Ibaraki, 305-0047, Japan*

Photoemission spectroscopy (PES) studies of $Sr_3Ni_2O_5Cl_2$ (Sr3252), a candidate ambient-pressure bilayer nickelate superconductor, have been performed to investigate the electronic structure. Valence band spectra, which are compared with DFT and DFT+$U$ calculations, reveal dominant Ni 3$d$ $t_{2g}$-derived states located around 0.6 eV binding energy with a shoulder around 0.2 eV that can be attributed to states with a dominant Ni 3$d_{3z^2-r^2}$ character. The overall valence band shape is better reproduced by the simulated spectrum of DFT+$U$ (= 3 eV), providing an experimental estimation of the strength of electron correlation. High-resolution PES near the Fermi level ($E_F$) exhibits marked temperature dependence, which cannot be fully understood by a broadening due to coupling of a photo-hole to phonons alone and suggests change in electronic structure. The results provide fundamental electronic structure information of Sr3252, which is useful to further explore the superconductivity of Sr3252.

## 1. Introduction

After the discovery of superconductivity in an infinite-layer nickelate [1], followed by the enhanced critical temperature ($T_c$) as high as 80 K in the bilayer nickelate $La_3Ni_2O_7$ (La327) under pressure [2], extensive studies have been underway to understand the mechanism of the superconductivity and to discover nickelates exhibiting higher $T_c$ values and to develop superconductors that exhibit high-$T_c$ superconductivity under ambient pressure [3].

Regarding the search for ambient-pressure high-$T_c$ nickelate superconductors, studies on thin films are a solid strategy, and indeed an increase in $T_c$ to 40 K has been reported [4]. Realization of ambient-pressure high-$T_c$ superconductivity in a bulk material may require new materials, as superconductivity seems tightly related to the tetragonal crystal structure [5], which cannot be realized under ambient pressure in La327 and $La_4Ni_3O_{10}$ (La4310)[6,7]. So far, theoretical studies have predicted several candidate superconductors that take a tetragonal structure at ambient pressure [8,9]. Interestingly, Ochi *et al.* theoretically predicted that $Sr_3Ni_2O_5Cl_2$ (Sr3252), which shares the same Ruddlesden-Popper-type crystal structure as La327, adopts a tetragonal structure under ambient pressure. They also predicted superconductivity with a $T_c$ exceeding that of La327 [9], based on the fluctuation exchange theory that explains superconductivity of La327 [10] and La4310 [6]. A possibility of spin-fluctuation-mediated superconductivity in La327 has been discussed before the discovery [11].

Motivated by this theoretical prediction, Yamane *et al.* successfully synthesized micro single crystals and confirmed the realization of a tetragonal crystal structure under ambient pressure [12]. Nevertheless, the temperature dependence of electrical resistivity exhibits a semiconductive behavior with a weak kink around $T$ = 200 K, and no superconductivity has been observed down to 2 K to date. Experimentally elucidating the electronic state of Sr3252 and comparing it with theoretical predictions is important to understand why this material does not exhibit metallic conducting properties and to understand what is necessary for the emergence of superconductivity.

In this work, we report photoemission spectroscopy studies of Sr3252, a candidate of ambient-pressure nickelate superconductors, to experimentally elucidate the electronic state. By comparing with theoretical calculations, we experimentally show the electron correlation effect, as well as the orbital characters of the valence band. We also observed a marked temperature dependence of the structure near the Fermi level ($E_F$), which gives insight into the semiconductive temperature dependence of electrical resistivity.

## 2. Experimental

The Sr3253 samples were synthesized via high-pressure techniques using a stoichiometric mixture of $SrO_2$, SrO, $SrCl_2$, and NiO. The detailed explanation will be found in Ref. [12]. Vacuum ultraviolet (VUV) photoemission spectroscopy (PES) experiments were performed at BL-28A, Photon Factory, with a Scienta DA30 spectrometer. The circularly polarized photons of $h\nu = 75$ eV with 18 meV total resolutions were used. The sample temperature was set to $T = 50$ K to reduce temperature broadening. The pressure inside the measurement chamber was below $1.6 \times 10^{-8}$ Torr. Hard x-ray PES (HAXPES) experiments were performed at BL09XU, SPring-8, with a Scienta R4000 spectrometer. The linearly polarized photons of $h\nu = 7940$ eV with a 180 meV total resolution were used. The sample temperature was set to $T = 20$ K. The pressure inside the measurement chamber was below $1.7 \times 10^{-7}$ Pa. High-resolution PES experiments were performed at Okayama Univ. with a Scienta R4000 spectrometer and the Xe Iα ($h\nu = 8.44$ eV) resonance line. The total energy resolution was set to 8 meV to compromise the extremely low intensity near $E_F$. The sample temperatures were set from $T = 10$ K to 300 K. The pressure inside the measurement chamber was below $5.1 \times 10^{-9}$ Torr. For all PES experiments, the clean sample surfaces were obtained by scraping the sample. The consistent spectra of 10 K before and after the measurement ensure negligible degradation during the measurement, which will be shown below.

First-principles band calculations were carried out by using the Perdew-Burke-Ernzerhof parametrization of the generalized gradient approximation [13] and the projector augmented-wave (PAW) method as implemented in *Vienna ab initio simulation package* (VASP)[14-18]. We used the experimental crystal structure reported in [12]. For the DFT+$U$ calculations, we included the +$U$ correction for Ni-d electrons with the simplified rotationally invariant approach [19]. We set a plane-wave cutoff energy of 600 eV for Kohn-Sham orbitals, and the spin-orbit coupling is not included for simplicity. In the self-consistent calculations, the Brillouin zone was sampled by a $12 \times 12 \times 12$ $k$ mesh. For calculating the density of states (DOS), we used a $22 \times 22 \times 22$ $k$ mesh.

## 3. Results and discussion

Figure 1 (a) shows the valence band spectra obtained with photon energies of $h\nu = 75$ eV and 7940 eV, normalized with the intensity at 4.5 eV binding energy. The spectrum of $h\nu = 75$ eV has four structures around 0.5, 2.0, 3.0, and 4.5 eV. The sharp peak around 0.5 eV is accompanied by a shoulder structure around 0.2 eV. We labeled them A, B, C, D, and E, as shown in Fig. 1(a). Near $E_F$, the intensity reduces toward $E_F$ and the intensity at $E_F$ is extremely small. The spectrum measured with the photon energy of $h\nu = 7940$ eV shows the highest intensity at 5.5 eV binding energy, which we labeled F. It is well known that spectra measured with HAXPES differ significantly in shape from spectra measured with low-energy photons, and this is explained by an increase in the photoionization cross-section of s and p orbital electrons in the hard x-ray region [20]. In Sr3252, the large enhancement of the structure F in the spectrum at $h\nu = 7940$ eV can be explained by the relatively larger photoionization cross section of Ni 4*s*, which shows a dominant structure at 5.5 eV binding energy (Fig.1(b)). For the lower binding energy region, the energy positions of the structures A-E and very small intensity at $E_F$, observed with $h\nu = 75$ eV photons, are similarly observed in the spectrum with $h\nu = 7940$ eV photons, ensuring that the spectrum at $h\nu = 75$ eV reflects bulk electronic states.

In Fig. 1(b), we show simulated spectra based on DFT and DFT+$U$ ($U$ = 2, 3, and 4 eV) results for comparison with the spectrum at $h\nu = 75$ eV. Calculated partial density of states (pDOS) of Ni *d*, O *p*, and Cl *p*, which give dominant contributions to the valence band for DFT and DFT+$U$ ($U$ = 3 eV), are also shown. The simulated spectra were obtained by summing pDOS of Ni *d*, O *p*, and Cl *p*, taking the photoionization cross sections of $h\nu = 75$ eV photons [21] into consideration, and were multiplied by the Fermi-Dirac distribution function of $T = 50$ K. With ease of comparison between the experimental spectrum and calculations, simulated spectra were tentatively broadened with Gaussian of 0.2 eV full width at half maximum. The simulated spectrum of DFT shows a high-intensity region from $E_F$ to 1 eV with a tendency of increasing intensity toward $E_F$. It also has a peak around 1.4 eV with a clear dip structure. While the peak with the dip structure looks correspond to the observed structure C, but the energy position of C is 1.7 eV, slightly different from that of the calculations. Moreover, the calculated structure near $E_F$ could not explain the experimental spectral shape near $E_F$ with the peak at 0.5 eV and decreasing intensity toward $E_F$. In the simulated spectrum based on DFT + $U$, the inclusion of $U$ systematically shifts the center of the gravity of the near $E_F$ structure to the higher binding energy side, compared to that of DFT. It also slightly increases the spectral weight around 4 eV. The overall spectral features of the simulated spectrum, especially for the structures corresponding to structures A, B, and C, of DFT+$U$ (= 3 eV) appear to better reproduce the experimental spectrum. From the $U$-induced higher binding energy shifts of the two structures corresponding to structures A and B, we may be able to set the region of $U$ value to 3 ±1 eV. The $U$ of 3.0 eV is in the same range of that reported for other nickelate La327 (3.5 eV) [22], giving a justification of the estimate of $U$ for Sr3252. This supports the use of $U = 3$ eV of theoretical studies which predicted superconductivity of Sr3252 [9].

The fact that the simulation better reproduces the experimental spectrum enables the assignment of the orbital origin of the structure. The structures near $E_F$ (around 4 eV) in the simulation, which are the hybridized states of Ni 3d and O 2p with an anti-bonding character (a bonding character) , correspond to the observed structures A and B (C). The states in between, corresponding to C and especially D, have larger contributions from Cl 3p orbitals with a non-bonding character. To investigate the Ni 3 *d* orbital components of the states near $E_F$ more in detail, we further compare the experimental spectrum with the orbital-resolved Ni 3*d* pDOS and O 2*p* pDOS from DFT + $U$ (= 3 eV), as shown in Fig. 2. The calculated states located between 0.9 and 0.1 eV and between -0.9 and -1.7 eV are bonding and anti-bonding components of $3d_{3z^2-r^2}$ bands, respectively. The $3d_{xy}$ band is located between 1.2 and 0.3 eV. The $3d_{yz}$ and $3d_{zx}$ bands are located between 0.9 and 0.1 eV. Since the Ni $d_{x^2-y^2}$-derived bands have a wider bandwidth and mostly is located above $E_F$, their intensity is small below $E_F$ and is rather flat above $E_F$. It is found that the energy position of the main peak of the simulation coincides with that of the experimental spectrum. The structure B is mainly derived from Ni 3*d* $t_{2g}$ band ($d_{xy}$, $d_{yz}$, and $d_{zx}$ orbitals) hybridized with O 2*p*. In addition, we find that the shoulder structure A of the experimental spectrum can be associated with the peak around 0.2 eV of the simulation that has a substantial $3d_{3z^2-r^2}$ character. It is reasonable to consider that the shoulder structure A is derived from the Ni $3d_{3z^2-r^2}$ orbital.

Regarding the theory for the high-$T_c$ superconductivity of nickelates, the Ni $3d_{3z^2-r^2}$ orbital plays a crucial role as it induces the spin fluctuation mediating Cooper pairing [9,10]. In La327, the binding energy of the Ni $3d_{3z^2-r^2}$ is reported to be around 50 meV binding energy. The deeper binding energy of the Ni $3d_{3z^2-r^2}$ orbital in Sr3252 than in La327 is due to combined effects of difference in the splitting energy of Ni $3d_{3z^2-r^2}$ bonding and antibonding bands, the energy separation between Ni $3d_{3z^2-r^2}$ and $3d_{x^2-y^2}$ levels, and the *d* electron number. But, good correspondence between the experimental spectrum and the calculation supports the estimation of larger electron occupation (close to half filled) of Ni $3d_{3z^2-r^2}$ in Sr3252 than in La327, which also supports the theoretical prediction of the superconductivity of Sr3252 [9].

It is interesting to compare the overall valence band structure of Sr3252 with that of La327 [24]. The valence band of La327 studied by HAXPES shows a broader band from 2 to 7 eV, which is attributed to hybridized states with O 2*p* and Ni 3*d* with a large contribution of La 5*p* due to enhanced photoionization cross section in the high-energy region. The spectrum has a structure around 1.5 eV, which is attributed mainly to Ni 3*d* $t_{2g}$ derived states. The intensity of the Ni $3d$ $e_g$ states near $E_F$ is negligibly small, which was discussed in relation to the density wave and/or its fluctuation. The Ni $3d$ $t_{2g}$ derived peak at 1.5 eV in La327 corresponds to the peak at 0.5 eV in Sr3252. The shift of the dominant Ni $3d$ band to $E_F$ in Sr3252 ($Ni^{3+}$) as compared to La327 ($Ni^{2.5+}$) is consistent with the electron doping in Sr3252 as is expected from the formal valence of Ni.

The observed overall valence band structures are reasonably well explained by the theoretical valence band structures that predict a metallic ground state and high-$T_c$ superconductivity [9]. In contrast, Sr3252 is experimentally non-metallic and non-superconductive [12]. To explore this discrepancy, we focus on the electronic structure near $E_F$ using a Xe resonance line, as shown in Fig. 3. In Fig. 3(a), the spectrum in the lower part shows structures around 0.5 eV and 0.15 eV binding energies. The structure around 0.5 eV corresponds to the structure B in Fig. 1(a) and is attributed to Ni $3d$ derived states. The structure around 0.15 eV is related to the structure A. The very small intensity at $E_F$ is consistent with the spectra measured at VUV and hard x-ray. In the upper part, spectra at $T$ = 10 K and 300 K are shown, which were normalized by the spectral weight between 0.45 and - 0.3 eV. The data exhibits large broadening of the spectral shape at $T$ = 300 K compared to that at $T$ = 10 K, which will be discussed in detail later.

To study the detailed temperature dependence of the electronic structure, we have performed photoemission experiments with a better energy resolution of 8 meV, as shown in the lower spectra in Fig. 3(b). The normalization of the spectra was made with the intensity at 0.17 eV binding energy, which is the cross point of the $T$ dependent normalized spectra in Fig. 3(a). The upper spectrum is the spectrum at each temperature divided by the Fermi-Dirac function convoluted with the Gaussian function corresponding to the total-energy resolution of each temperature. As the temperature is lowered, the spectral intensity around 0.15 eV evolves with the reduction of intensity at $E_F$. At first glance, this change might seem to indicate pseudogap formation. However, considering the significant broadening of the spectra (Fig. 3 (a)) in the wider energy range, further analyses are needed.

To investigate the effect of broadening, we followed the previous studies, where temperature dependent spectra were analyzed by gaussian function that representing the coupling of a photo-hole to phonons. This analysis was first applied to core level spectra [25], but it was also applied to valence bands [26,27]. In the present study, a model DOS was created based on the lowest temperature data at $T$ = 10 K (a dotted line in Fig. 4). We assumed a constant DOS above $E_F$ that smoothly connects to the spectrum. The model DOS was convolved with a Gaussian function with a different FWHM. The Fermi distribution function of $T$ = 300 K was then applied to obtain a simulation spectrum. The simulated spectra roughly reproduce the spectral shape of the experimental spectrum at $T$ = 300 K with FWHM of 150 meV. Note that the phonon broadening of 150 meV is not unreasonably large, as the valence band broadening of about 200 meV at 300 K was reported for gold [28]. Near $E_F$ (inset), the larger the FWHM is, the higher the intensity at $E_F$ becomes. This indicates that $T$-dependent broadening itself can increase the intensity at $E_F$, as observed, which prevents us from concluding pseudogap formation at lower $T$. On the other hand, it is found that, although simulated spectrum with FWHM of 150 meV can reproduce the higher binding energy side of the experimental spectrum, it is difficult to reproduce the spectral shape around 70 meV by changing FWHM values (inset of Fig. 4). This indicates that the observed $T$ dependence cannot be explained by the broadening alone and suggests change in the electronic structure itself in this $T$ range. In La327, an anomaly in the temperature-dependent resistivity data at ambient pressure has been discussed in terms of electronic ordering (charge and spin density waves) [29]. The relation of this order to the high-$T_c$ superconductivity in nickelates is one of the open questions. In Sr3252, $T$-dependent resistivity data exhibit a weak kink around $T$ = 200 K. This, combined with the present study, might suggest that in Sr3252 some kind of ordered phase might form and obstacle superconductivity. If this is the case, preventing formation of the order is crucial to achieve superconductivity. However, at present, limited data on the physical properties of this material make it difficult to further discuss the origin of the electronic structure change. Understanding the origin may be an important step toward the realization of the superconductivity of Sr3252.

## 4. Summary

In summary, we have performed photoemission spectroscopy studies of Sr3252, which is a candidate high-$T_c$ superconductor under ambient pressure. Valence band spectra show that dominant Ni 3$d$ derived states are located around 0.6 eV binding energy with a shoulder around 0.2 eV that can be attributed to states with a Ni $3d_{3z^2-r^2}$ character. The overall valence band shape is better reproduced by the simulated spectrum of DFT+$U$ (= 3 eV), giving the experimental value of the electron correlation $U$. High-resolution PES near $E_F$ exhibits the large $T$-dependent change in the spectral shape. The spectral change cannot be explained by the Gaussian function broadening alone, suggesting a change in the electronic structure possibly due to unknown electronic orderings. These results constitute fundamental electronic structure information of Sr3252 and are also useful to consider strategies to realize metallization and superconductivity of Sr3252.

**Acknowledgment**

We thank Nae Yamaguchi, Mayuka Kumura, Shotaro Ikeda, and Atsuhiro Matsuno for their support during photoemission experiments. The PES experiments at PF were performed under the Photon Factory Proposal Nos. 2024U001 and 2026G126. The PES experiments at SPring-8 were performed under proposal numbers 2025A1739 and 2025B1917. This work was partially supported by JSPS KAKENHI Grant Numbers (JP23H0112515, JP25K07184, JP25K04857, JP26H02014) from MEXT, JST K Program Grant Number JPMJKP25Z3, and JSPS program for Forming Japan's Peak Research Universities (J-PEAKS) Grant Number JPJS00420230010.

*E-mail: yokoya@cc.okayama-u.ac.jp

1) D. Li *et al.*, Nature **572**, 624 (2019).
2) H. Sun *et al.*, Nature **621**, 493 (2023).
3) As a review, see Y. Wang *et al.*, National Science Review **12**, nwaf373 (2025).
4) G. Zhou et al., Nature **640**, 641-646 (2025).
5) L. Wang et al., J. Am. Chem. Soc. **146**, 7506-7514 (2024).
6) H. Sakakibara et al., Phys. Rev. B **109,** 144511 (2024).
7) Y. H. Zhu et al., Narure **631**, 531 (2024).
8) S. Wu *et al.*, arXiv: 2403.11713.
9) M. Ochi *et al.*, Phys. Rev. B **111,** 064511 (2024).
10) H. Sakakibara *et al.* Phys. Rev. Lett. **132**, 106002 (2024).
11) M. Nakata, D. Ogura, H. Usui, and K. Kuroki, Phys. Rev. B **95**, 214509 (2017).
12) K. Yamane *et al.*, Acta. Cryst. **C81**, 259 (2025).
13) J. P. Perdew *et al*, Phys. Rev. Lett. **77**, 3865 (1996)
14) G. Kresse and D. Joubert, Phys. Rev. B **59**, 1758 (1999).
15) G. Kresse and J. Hafner, Phys. Rev. B **47**, 558 (1993).
16) G. Kresse and J. Hafner, Phys. Rev. B **49**, 14251 (1994).
17) G. Kresse and J. Furthmüller, Comput. Mater. Sci. **6**, 15 (1996).
18) G. Kresse and J. Furthmüller, Phys. Rev. B **54**, 11169 (1996).
19) S. L. Dudarev *et al.*, Phys. Rev. B **57**, 1505 (1998).
20) D. Takegami *et al.*, Phys. Rev. B **99**, 165101 (2019).
21) J. J. Yeh and I. Lidau, Atomic data and nuclear data Tables **32**, 1-155 (1985).

22) Jiangang Yang *et al.,* Nature **15**, 4373 (2024).
23) H. Sakakibara *et al*. Phys. Rev. Lett. **132**, 106002 (2024).
24) D. Takegami *et al.*, Phys. Rev. B **109**, 125119 (2024).
25) J. A. D. Matthews and M. G. Devey, J. Phys. C. **7**, L335 (1974).
26) G. K. Wertheim, J. E. Rowe, D. N. E. Buchanan, and P. H. Citrin, Phys. Rev. B **51**, 13675 (1995).
27) G. K. Wertheim, Phys. Rev. B **51**, 10248 (1995).
28) T. Yokoya, A. Chainani, and T. Takahashi, Phys. Rev. B **55**, 5574 (1997).
29) Z. Liu *et al.*, Sci. China-Phys. Mech. Astron. **66**, 217411 (2023).

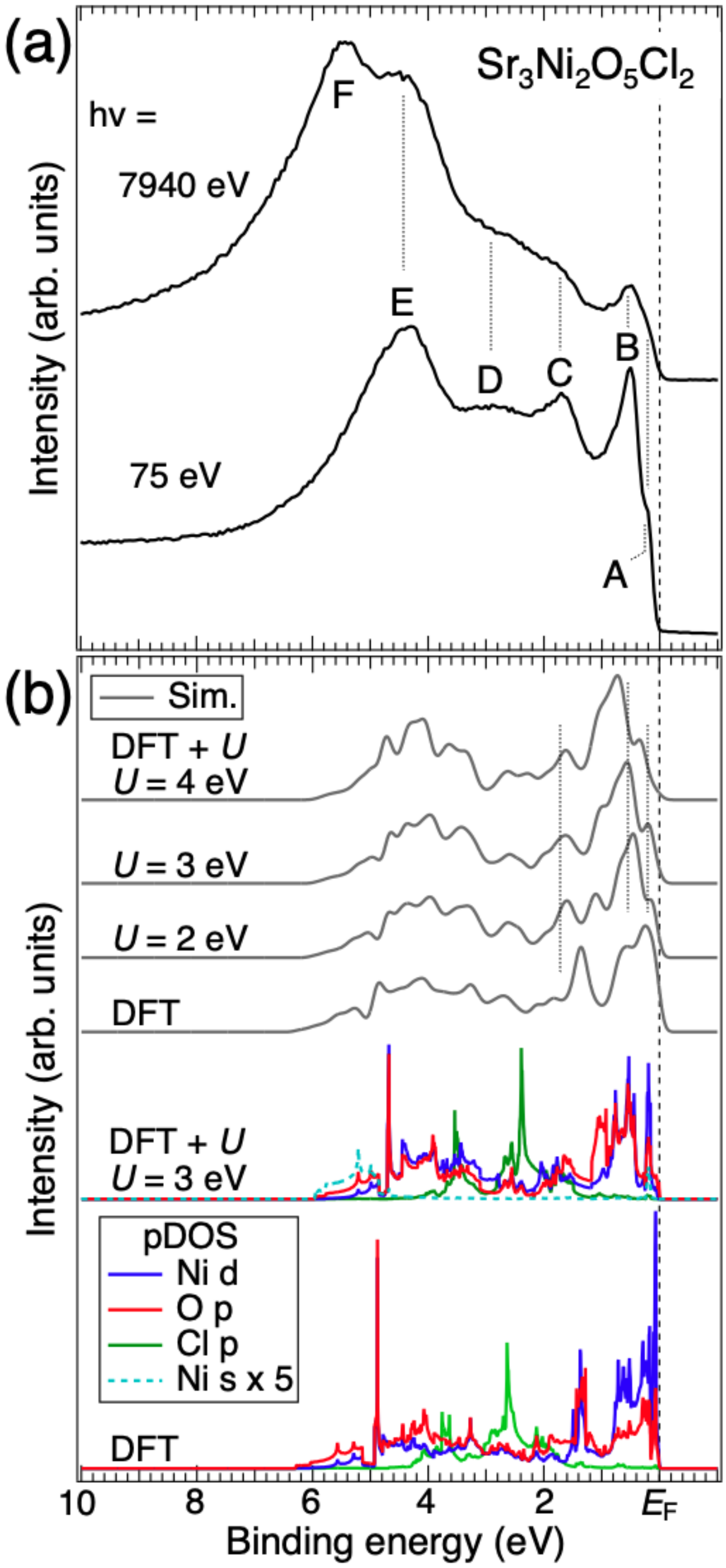

Fig. 1. (a) The valence band spectra obtained with the photon energies of $h\nu$ = 75 and 7940 eV, normalized with the intensity at 4.5 eV binding energy. (b) Simulated spectra based on DFT and DFT + $U$ (=2, 3, and 4 eV) (gray lines). Partial DOS of Ni 3d, O 2p, and Cl 3p for DFT and DFT + $U$ ($U$ = 3 eV) are also shown. For DFT + $U$ ($U$ = 3 eV), Ni 3s is shown. The three vertical lines show the binding energies of the structures A, B, and C in (a).

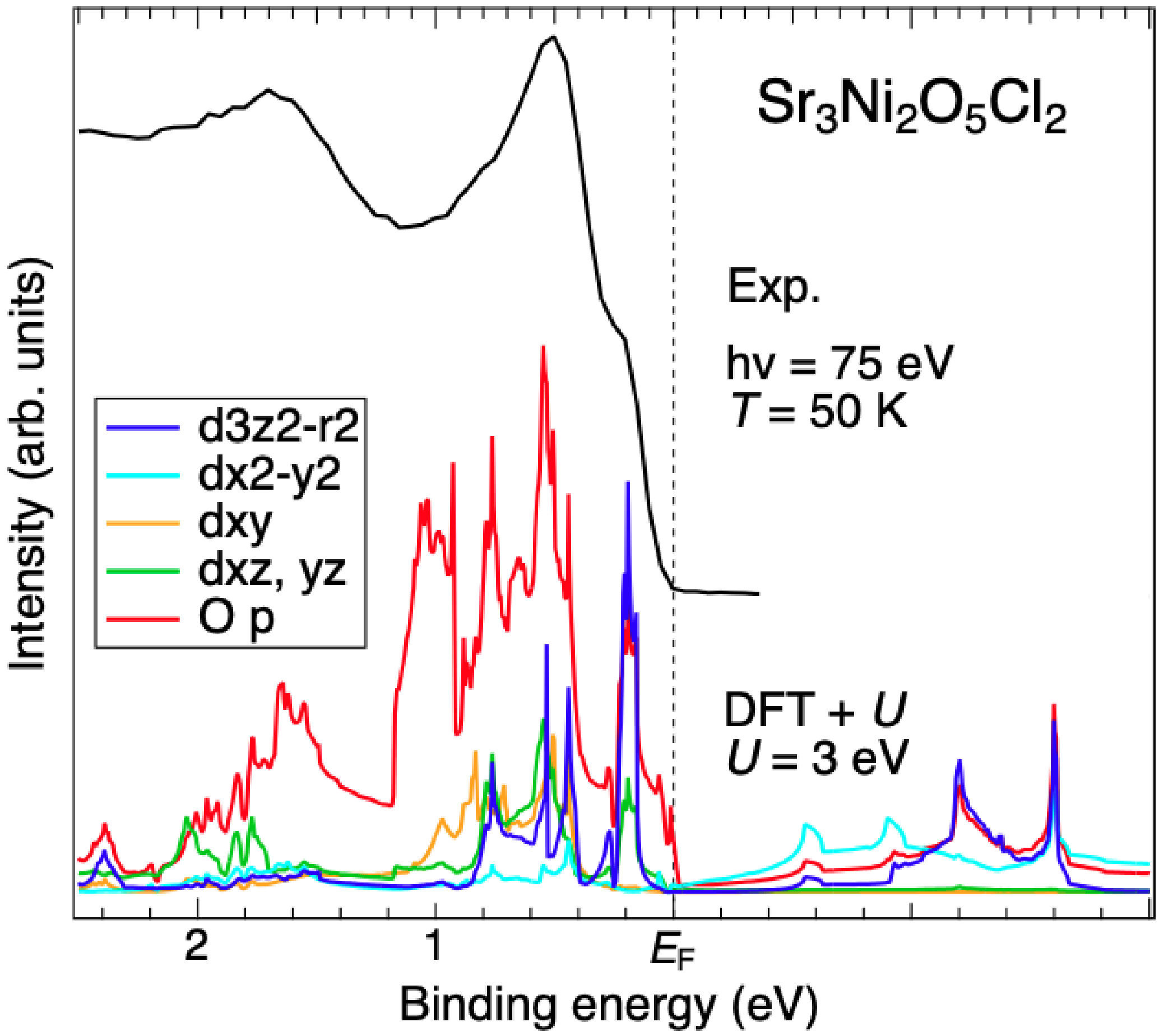

Fig. 2. The valence band spectrum measured at $h\nu$ = 75 eV, compared with partial DOS of $3z^2$-$r^2$, $x^2$-$y^2$, xy, xz, and yz of Ni 3d, as well as O 2p.

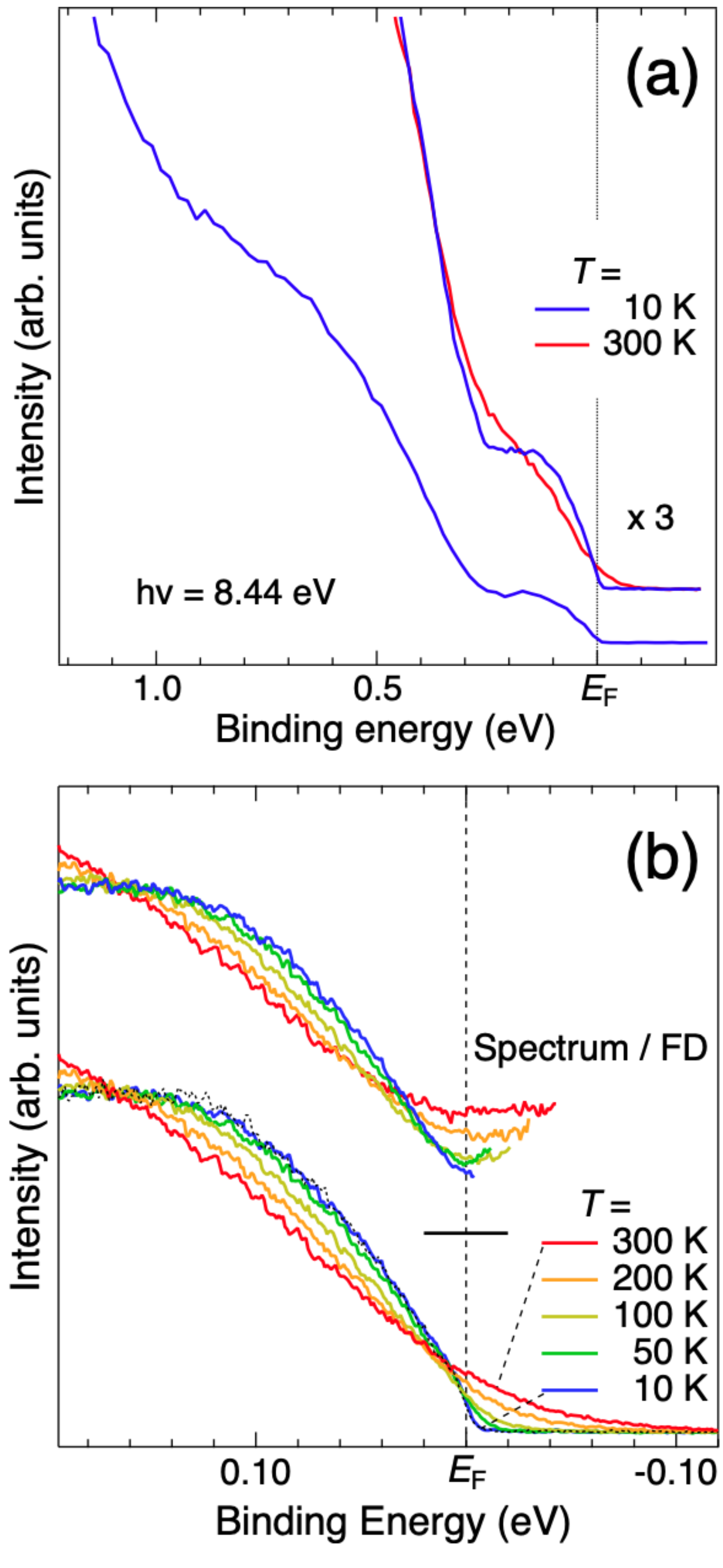

Fig. 3. (a) Photoemission spectrum of Sr3252 measured with a Xe resonance line and at $T = 10$ K. Intensity enlarged spectra near $E_F$ measured at $T = 10$ K and 300 K are also shown. (b) $T$-dependence of high-resolution photoemission spectra of Sr3252 (bottom) and the same data divided by Fermi-Dirac distribution functions of corresponding $T$ (top). The measurement was done in a sequence of 10 K, 50 K, 100 K, 200 K, 300 K and 10 K. The spectrum at 10 K after measuring at 300 K is superimposed on the raw data as a dotted line.

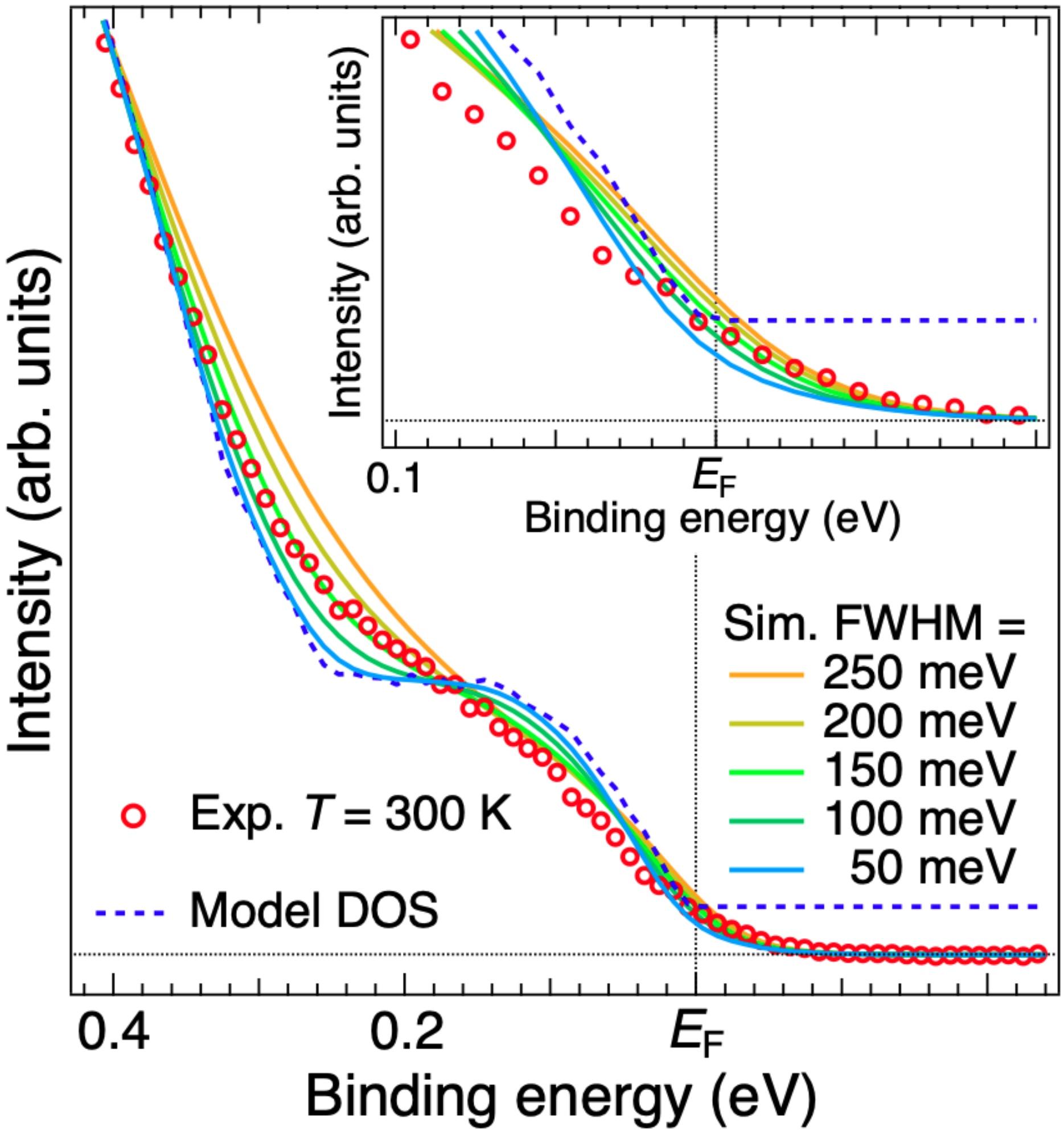

Fig. 4. Simulated spectra (lines) broadened with different values of FWHM of Gaussian functions compared with the experimental spectrum at $T$ = 300 K (open circles). Model DOS (broken line) is also shown. Inset: enlarged data of the near-$E_F$ region.